\documentclass[runningheads]{llncs}

\usepackage{graphicx}
\usepackage{xcolor}%

\usepackage{subcaption}
\usepackage{amsmath}
\usepackage[backend=biber, sorting=none]{biblatex}
\usepackage{comment}

\usepackage{multirow}
\usepackage{array}
\usepackage{lscape}
\usepackage{textcomp}
\usepackage{xspace}
\usepackage{hyperref}

\def\ECCVSubNumber{5}  

\let\OldTexttrademark\texttrademark
\renewcommand{\texttrademark}{\OldTexttrademark\xspace}%

\begin{document}

\title {\Large{\textbf{Learning with minimal effort: leveraging in silico labeling for cell and nucleus segmentation}}}

\author{Thomas Bonte\inst{1,2,3} \and
Maxence Philbert\inst{1,2,3} \and
Emeline Coleno\inst{4} \and
Edouard Bertrand\inst{4} \and
Arthur Imbert\inst{1,2,3,5} \and
Thomas Walter\inst{1,2,3,5}\orcidID{0000-0001-7419-7879}}
\authorrunning{T. Bonte et al.}
\titlerunning{Learning with minimal effort: leveraging ISL for segmentation}

\institute{Centre for Computational Biology (CBIO), Mines Paris, PSL University, 75006 Paris, France 
\email{\{Thomas.Bonte, Thomas.Walter\}@minesparis.psl.eu}
\and Institut Curie, PSL University, 75248 Paris Cedex, France 
\and INSERM, U900, 75248 Paris Cedex, France
\and IGH, University of Montpellier, CNRS, 34090 Montpellier, France
\and corresponding authors}

%
\maketitle              
\begin{abstract}
Deep learning provides us with powerful methods to perform nucleus or cell segmentation with unprecedented quality. However, these methods usually require large training sets of manually annotated images, which are tedious --- and expensive --- to generate. In this paper we propose to use In Silico Labeling (ISL) as a pretraining scheme for segmentation tasks. The strategy is to acquire label-free microscopy images (such as bright-field or phase contrast) along fluorescently labeled images (such as DAPI or CellMask\texttrademark). We then train a model to predict the fluorescently labeled images from the label-free microscopy images. By comparing segmentation performance across several training set sizes, we show that such a scheme can dramatically reduce the number of required annotations.

\keywords{Segmentation \and Transfer learning \and Pretext task \and In Silico Labeling
\and Fluorescence microscopy}
\end{abstract}

\section{Introduction}

Detection and segmentation of cells and nuclei, among other cell structures, are essential steps for microscopy image analysis. Deep Learning has provided us with very powerful methods to perform these segmentation tasks. In particular, recently published neural networks, such as NucleAIzer \cite{nucleaizer}, Cellpose \cite{cellpose} or StarDist \cite{stardist}, trained on hundreds of images of different modalities, give  excellent results, outperforming by far traditional methods for image segmentation. However, the main drawback of state-of-the-art networks is the need for large amounts of fully annotated ground truth images, which can take a significant amount of time to create. Here, we present an alternative strategy, where we pre-train our segmentation models using In Silico Labeling (ISL)  before fine-tuning them on a very small data set to perform nucleus and cell segmentation.  

ISL was first introduced by \cite{CHRISTIANSEN2018792}, aiming to predict fluorescent labels from bright-field inputs. Fluorescence microscopy is the major technique employed in cellular image-based assays, as the use of fluorescence labels allows to highlight particular structures or phenotypic cell states. However, the number of fluorescent labels is limited (typically up to 4). In addition, phototoxicity and photobleaching can also represent serious drawbacks. 

To tackle these limitations, several variants have been proposed. In \cite{Ounkomol2018}, ISL is applied to predict fluorescent labels from transmitted-light images (DIC), or immunofluorescence from electron micrographs. Besides, Generative Adversarial Networks (GAN) are used in \cite{Rivenson2019} to predict different stains: H\&E, Jones Silver or Masson's trichrome. They underlie staining standardization as an advantage of ISL. In another paper \cite{phasestain} GANs are also used on different transmitted light images: quantitative phase images (QPI). Moreover, in \cite{Li2020} conditional GANs (cGAN) generate H\&E, PSR and Orcein stained images from unstained bright-field inputs. In \cite{liu2020}, using the same data set and same tasks as \cite{CHRISTIANSEN2018792}, the authors add attention blocks to capture more information than usual convolutions. Finally, stained images of human sperm cells are generated in \cite{yoav2020}, from quantitative phase images. They use these virtually stained images to recognize normal from abnormal cells. The principle of ISL has also been proposed for experimental ground truth generation for training cell classifiers for the recognition of dead cells \cite{JBoyd2020,Hu2022}, tumour cells \cite{Zhang2022-dq} embryo polarization \cite{Shen2022} or the cell cycle phase \cite{popescuCycle}.

In this paper we show that models trained to generate fluorescence microscopy images with nuclear or cytoplasmic markers can be used efficiently to pretrain segmentation networks for nuclear and cell segmentation, respectively. To the best of our knowledge, no previous work has used ISL as a pretext task for segmentation of cell structures. This provides us with a powerful strategy to minimize the annotation burden for a given application, and to train models on large data sets, requiring only minimal effort in terms of manual annotation. 

\section{Materials and Methods}

\subsection{Image Acquisition}

We work on two different data sets. The first dataset has been generated by the Opera Phenix\texttrademark Plus High-Content Screening System (Perkin Elmer). It contains 960 images of dimension (2160, 2160). For each position, we acquired bright-field images and DAPI, both at 4 different focal planes. DAPI is a very common fluorescent stain binding to AT-rich regions of the DNA, which can thus be used to locate the nucleus in eukaryotic cells. Additionally we have a phase contrast image, computationally created from the 4 bright-field images by a proprietary algorithm of the Opera system. Images contain on average 15.6$\pm$19.6 cells.

Our second data set contains 100 images of dimension (1024, 1024). We used Differential Interference Contrast (DIC) as label-free microscopy technique, and we marked the cytoplasmic membrane with the CellMask\texttrademark marker (Life Technologies). Images contain on average 52.4$\pm$15.1 cells.

\subsection{Nucleus Segmentation}

Nucleus segmentation is one of the most important segmentation tasks in biology, as nuclear morphologies are indicative of cellular states, and because they are visually very different from the cytoplasm. Segmentation of the nucleus is usually a comparatively simple segmentation task, and for this reason we assumed that this might be a good first segmentation problem to investigate our ISL-based pretraining. 

\subsubsection{DAPI prediction as pretraining task}

The first step of our strategy for nucleus segmentation is the prediction of DAPI images from bright-field inputs.

\begin{figure}
\centering
\begin{subfigure}[b]{0.3\textwidth}
\centering
   \includegraphics[width=\linewidth]{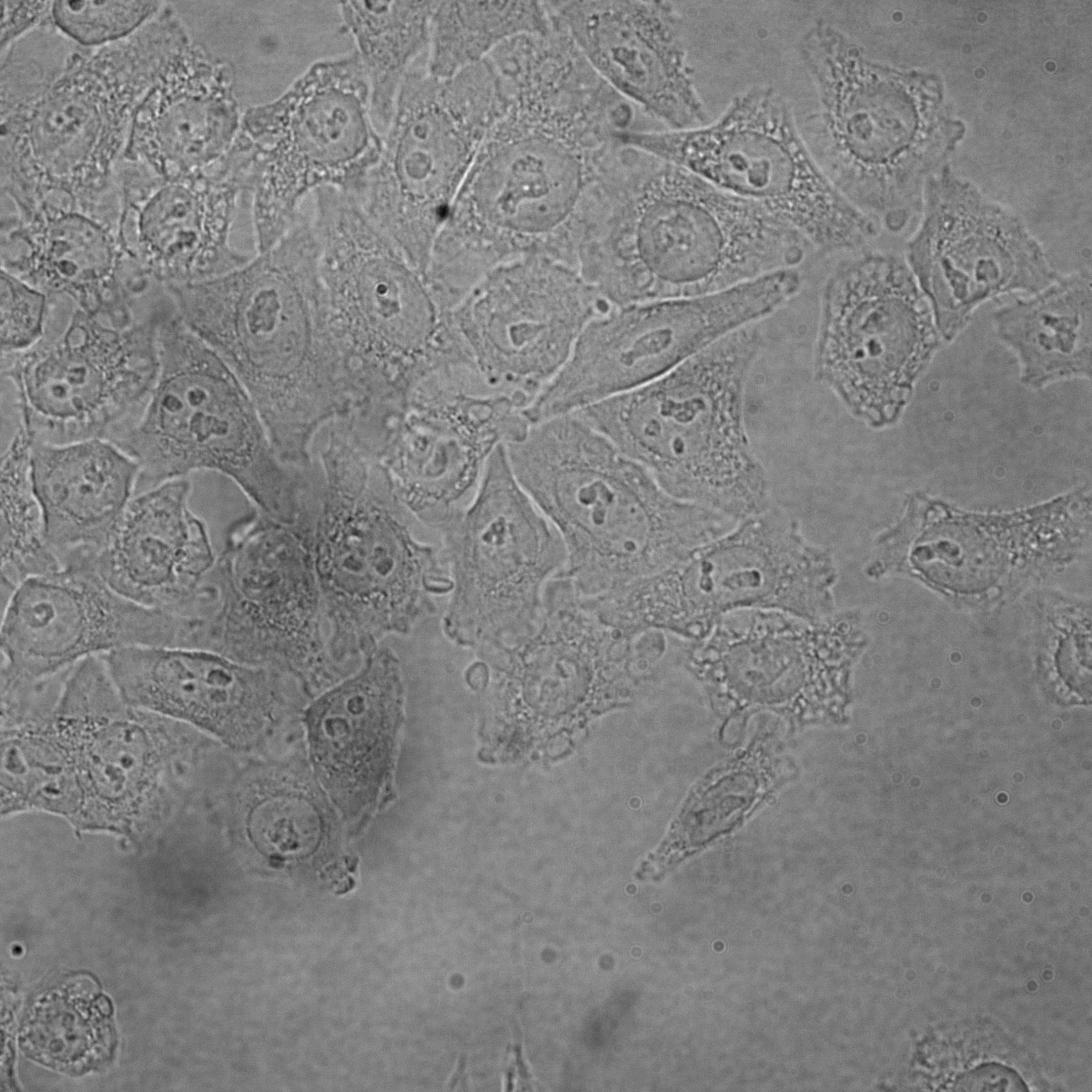}
   \caption{}
\end{subfigure}%
~
\begin{subfigure}[b]{0.3\textwidth}
\centering
   \includegraphics[width=\linewidth]{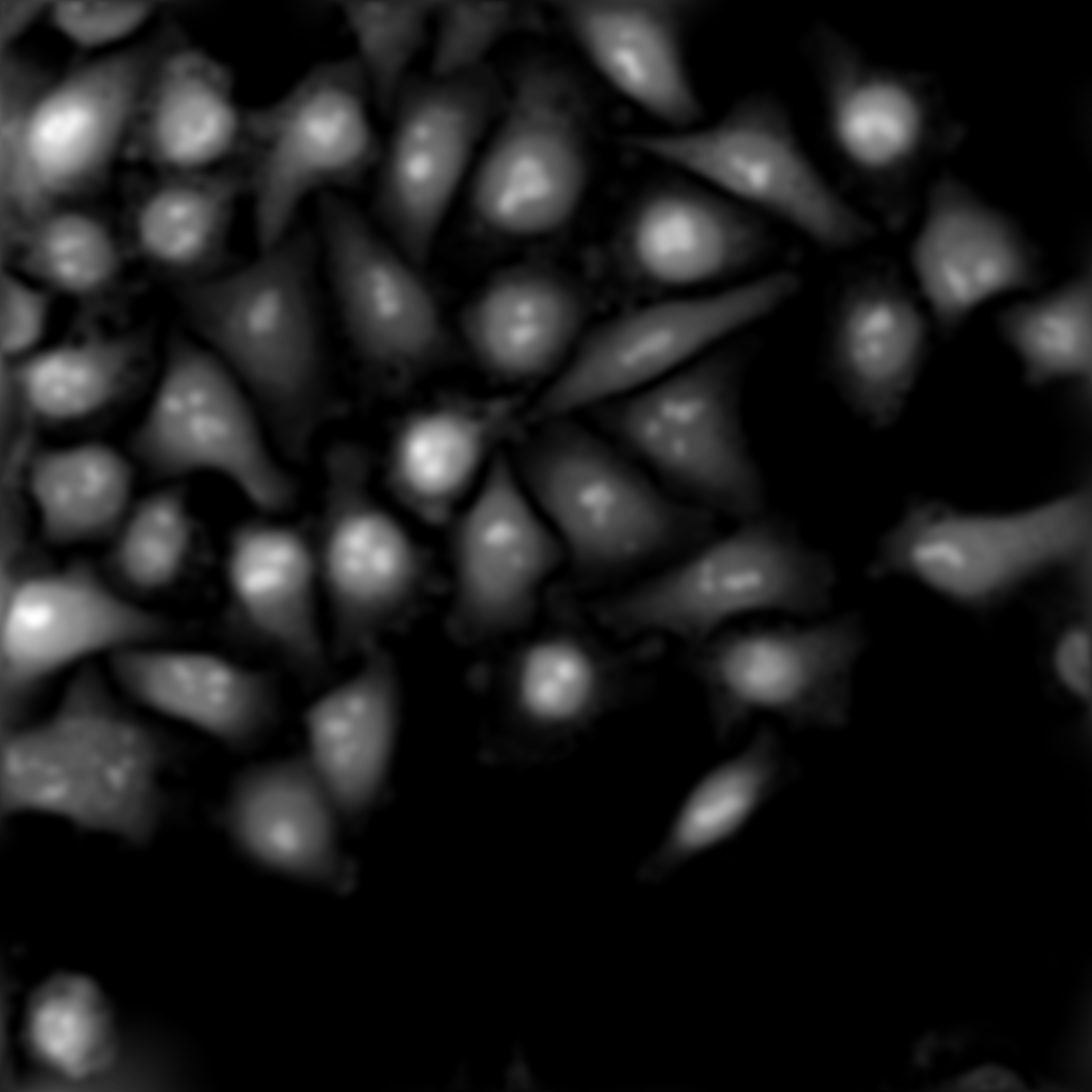}
   \caption{}
\end{subfigure}%
~
\begin{subfigure}[b]{0.3\textwidth}
\centering
   \includegraphics[width=\linewidth]{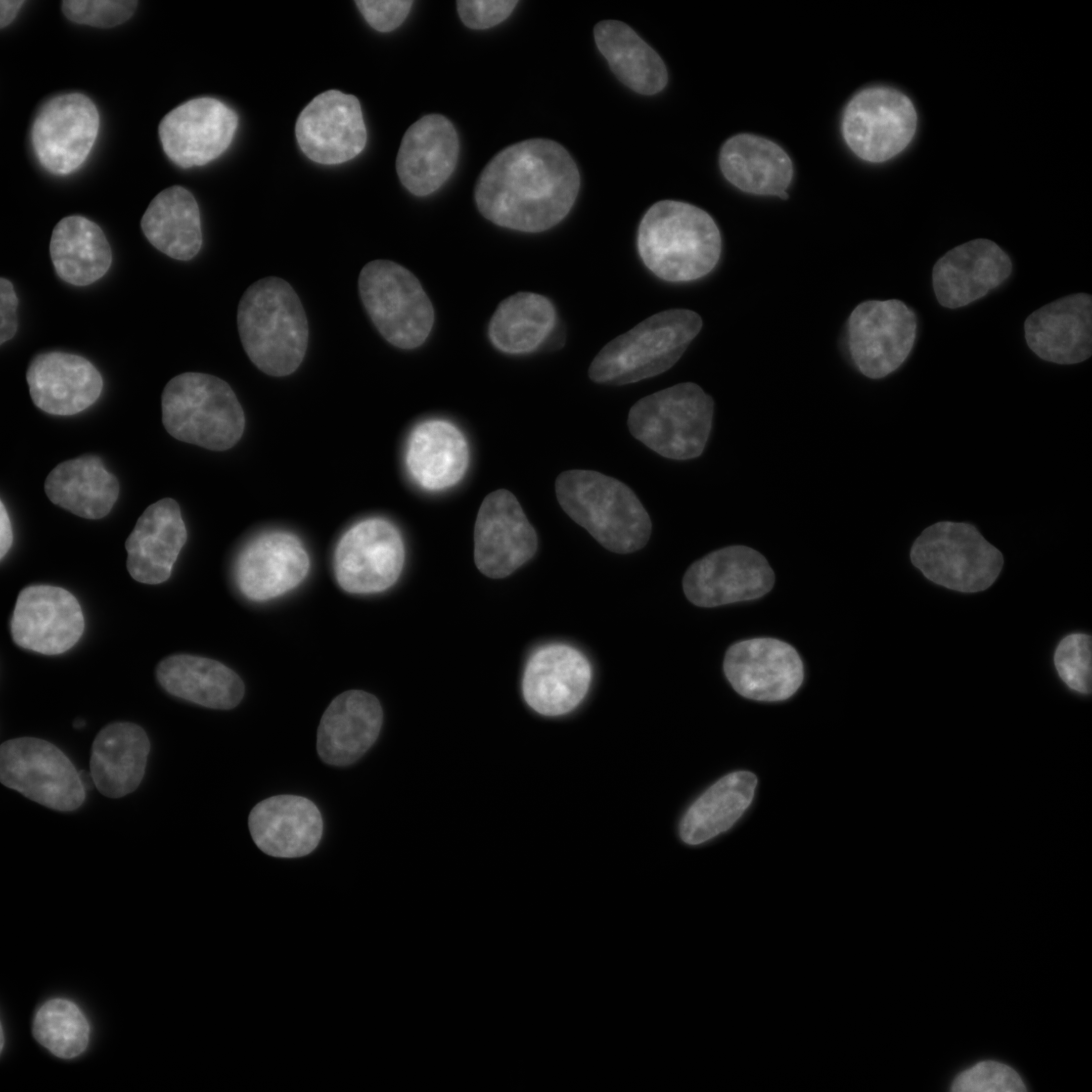}
   \caption{}
\end{subfigure}

\caption{Images from the same field-of-view, for a given focal plane. (a) Bright-field image. (b) Phase contrast image, computationally generated by the Opera system. (c) Fluorescent DAPI image.}
\label{fig:data_set_1}

\end{figure}

We used a data set of 421 images of dimension (2160, 2160), divided into 384 images for training and 37 images for testing. 5 images of dimension (512, 512) were randomly cropped from each initial image (see Fig.\ref{fig:data_set_1}). Note that we only included images containing at least one nucleus.

Inspired by the work of \cite{CHRISTIANSEN2018792}, the model is a U-net-shape model \cite{unetpaper} with a densenet121 architecture \cite{Yakubovskiy:2019}. It has been previously trained on ImageNet \cite{deng2009imagenet}, hence it is referred to as 'on steroids' in the following. As input we used 3 channels, 2 being bright-field images of the same field-of-view with different focal planes, and the third the corresponding phase-contrast image. As output we used only one channel, the maximum intensity projection of our DAPI images (z-stack, 4 focal planes) that we have for each field-of-view. 

We did not use any data augmentation. All training details are reported in Supplementary Table 1.

\subsubsection{Transfer Learning for Nucleus Segmentation.}

In a first step, we aimed at investigating how pretraining on fluorescent markers impacts semantic segmentation. For this, we turned to nucleus segmentation.  

In order to generate the ground truth, we applied Cellpose \cite{cellpose}, a widely used segmentation technique in bioimaging, based on a U-net-shaped network, trained on massive amounts of heterogeneous data. We applied Cellpose to the DAPI channel and corrected the segmentation results manually. As segmentation of nuclei from DAPI images with high resolution is a fairly simple task, as expected the results were overall excellent.

Next, we used training sets with different sizes $N \in \{1, 10, 50, 100, 200, 500\}$, composed of images of dimension (2160, 2160) and evaluated the accuracy for each $N$. Testing is always performed on the same withheld 190 images. 5 images of dimension (512, 512) were randomly cropped from each initial image.

\begin{figure}
\centering
\begin{subfigure}[b]{0.8\textwidth}
   \centering
   \includegraphics[width=1\linewidth]{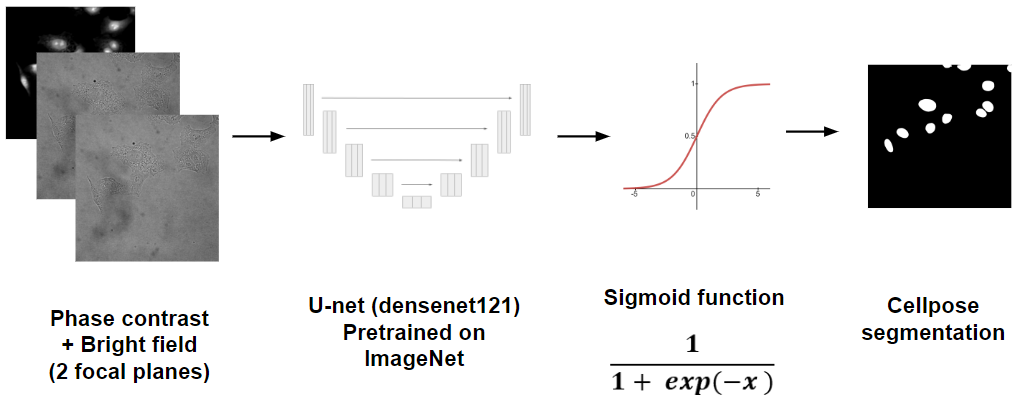}
   \caption{}
   \label{fig:nuclei_steroids}
\end{subfigure}

\begin{subfigure}[b]{0.8\textwidth}
    \centering
   \includegraphics[width=1\linewidth]{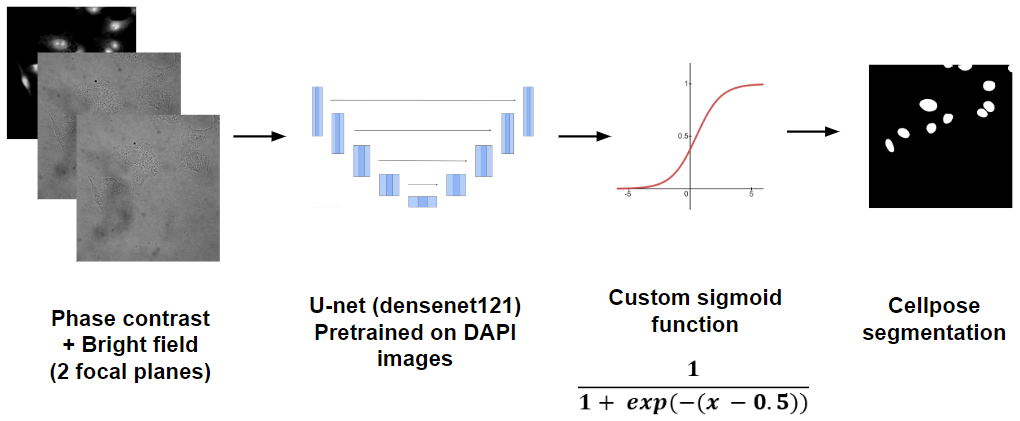}
   \caption{}
   \label{fig:nuclei_pretrained}

\end{subfigure}

\caption{Compared models to predict nucleus semantic segmentation. (a) U-net 'on steroids' which has not been trained on DAPI images. (b) U-net 'on steroids' pretrained on DAPI images. Note the difference in the activation functions.}
\end{figure}

To investigate whether our pretraining scheme is useful for segmentation, we compare two different models. The first model is composed of the U-net 'on steroids' followed by a sigmoid activation function in order to output, for each pixel, its probability of belonging to a nucleus (Fig.\ref{fig:nuclei_steroids}). The second model has the same U-net architecture but is pretrained on DAPI images, and has an activation function displayed in equation \eqref{customsigmoid} that takes a different range into account (Fig.\ref{fig:nuclei_pretrained}). The reason for this choice is that the model pretrained on DAPI images is likely to output values between 0 and 1, so we centered the following activation function around 0.5.

\begin{equation}
f(x) = \frac{1}{1 + \exp(-(x - 0.5))}
\label{customsigmoid}
\end{equation}

We did not use any data augmentation. All training details are reported in Supplementary Table 1.

\subsection{Cell Segmentation}

We next turned to the application of our pretraining scheme to cell segmentation, a more difficult multiple instance segmentation scenario. 

\subsubsection{CellMask\texttrademark Prediction as Pretraining Task}

\begin{figure}
\centering
\begin{subfigure}[b]{0.3\textwidth}
\centering
   \includegraphics[width=0.75\linewidth]{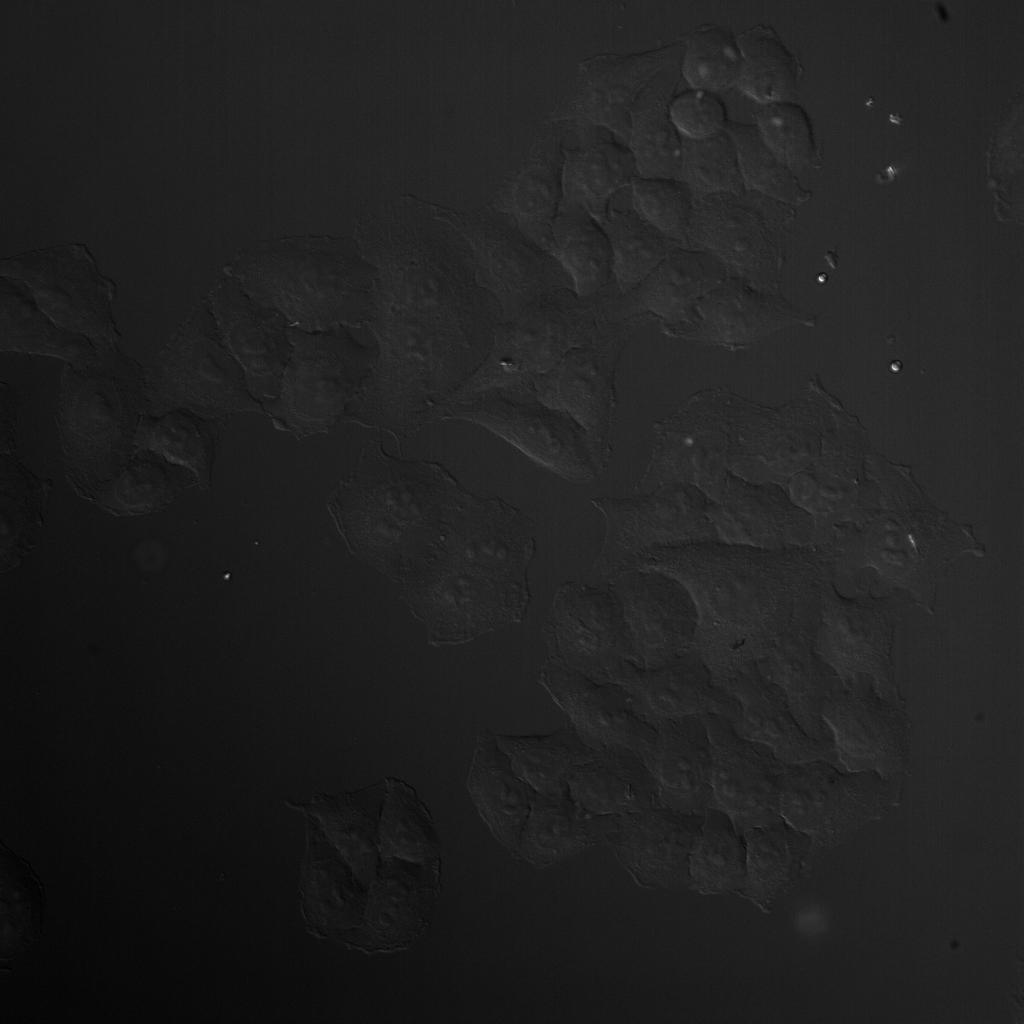}
   \caption{}
\label{fig:dic}
\end{subfigure}%
~
\begin{subfigure}[b]{0.3\textwidth}
\centering
   \includegraphics[width=0.75\linewidth]{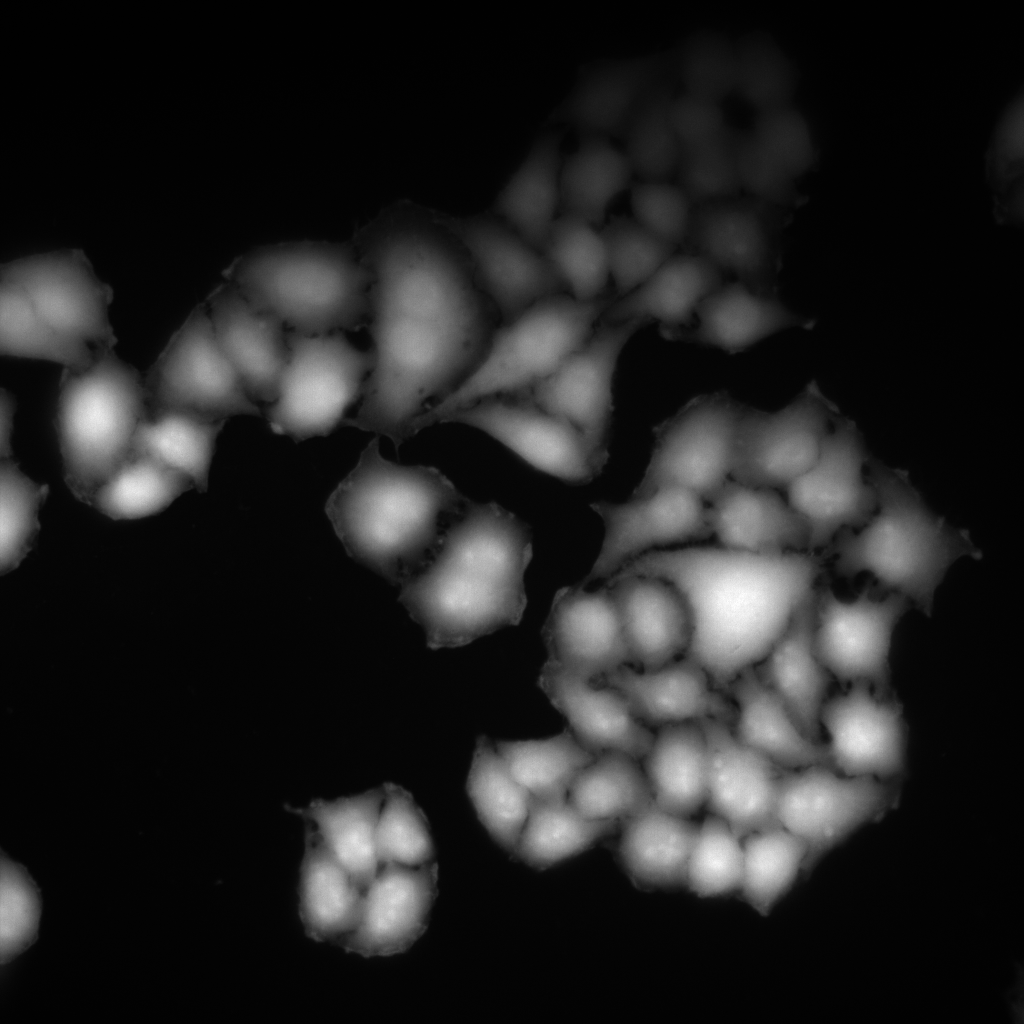}
   \caption{}
\label{fig:cy5}
\end{subfigure}

\caption{Images from the same field-of-view, for a given focal plane. (a) DIC image. (b) Fluorescent CellMask\texttrademark image.}
\end{figure}

In our pretraining strategy, the first step of cell segmentation is the prediction of CellMask\texttrademark (Fig.\ref{fig:cy5}) images from DIC microscopy as inputs (Fig.\ref{fig:dic}). 

We used a data set of 100 images of dimension (1024, 1024), divided into 90 images for training and 10 images for testing. 5 images of dimension (512, 512) were randomly cropped for each initial image.

For comparison, we again used the U-net 'on steroids'. We did not use any data augmentation. All training details are reported in Supplementary Table 2.

\subsubsection{Transfer Learning for Cell Segmentation.}

Segmentation of cells is usually more difficult than nuclear segmentation, because cells tend to touch each other, and the precise detection of the contact line can be challenging. Indeed, we need to turn to multiple instance segmentation, where object properties are predicted together with pixel labels. 

\begin{figure}
\centering
\begin{subfigure}[b]{0.2\textwidth}
\centering
   \includegraphics[width=1\linewidth]{images_V1/TWDIC2_w4Cy5_s1_z2.png}
   \caption{}
\label{fig:cy5_bis}
\end{subfigure}%
~
\begin{subfigure}[b]{0.2\textwidth}
\centering
   \includegraphics[width=1\linewidth]{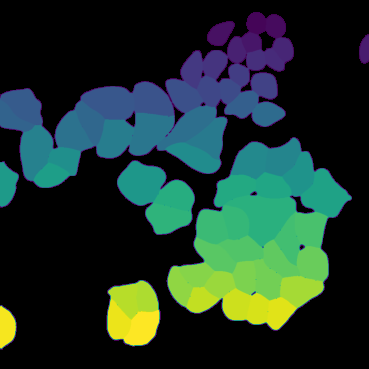}
   \caption{}
\label{fig:cell_instance}
\end{subfigure}%
~
\begin{subfigure}[b]{0.2\textwidth}
\centering
   \includegraphics[width=1\linewidth]{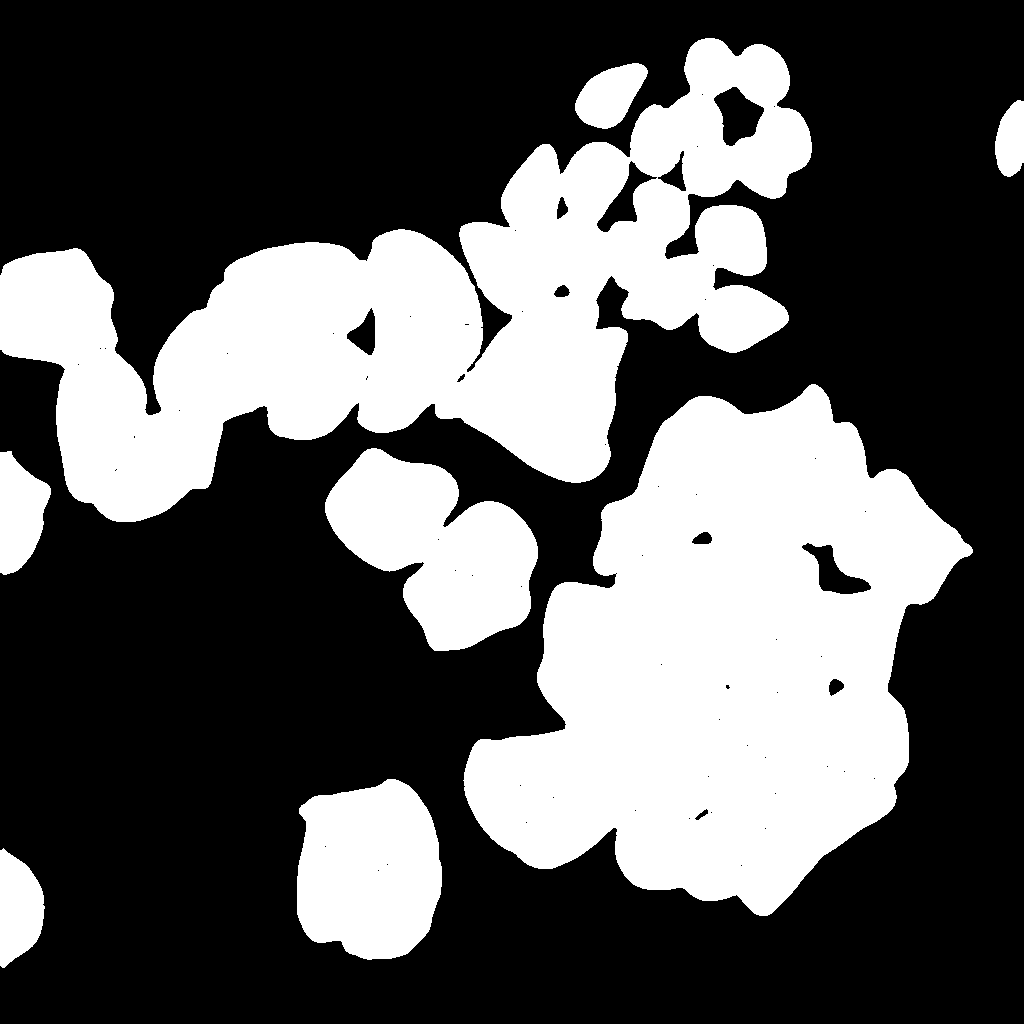}
   \caption{}
   \label{fig:cell_semantic}
\end{subfigure}%
~
\begin{subfigure}[b]{0.2\textwidth}
\centering
   \includegraphics[width=1\linewidth]{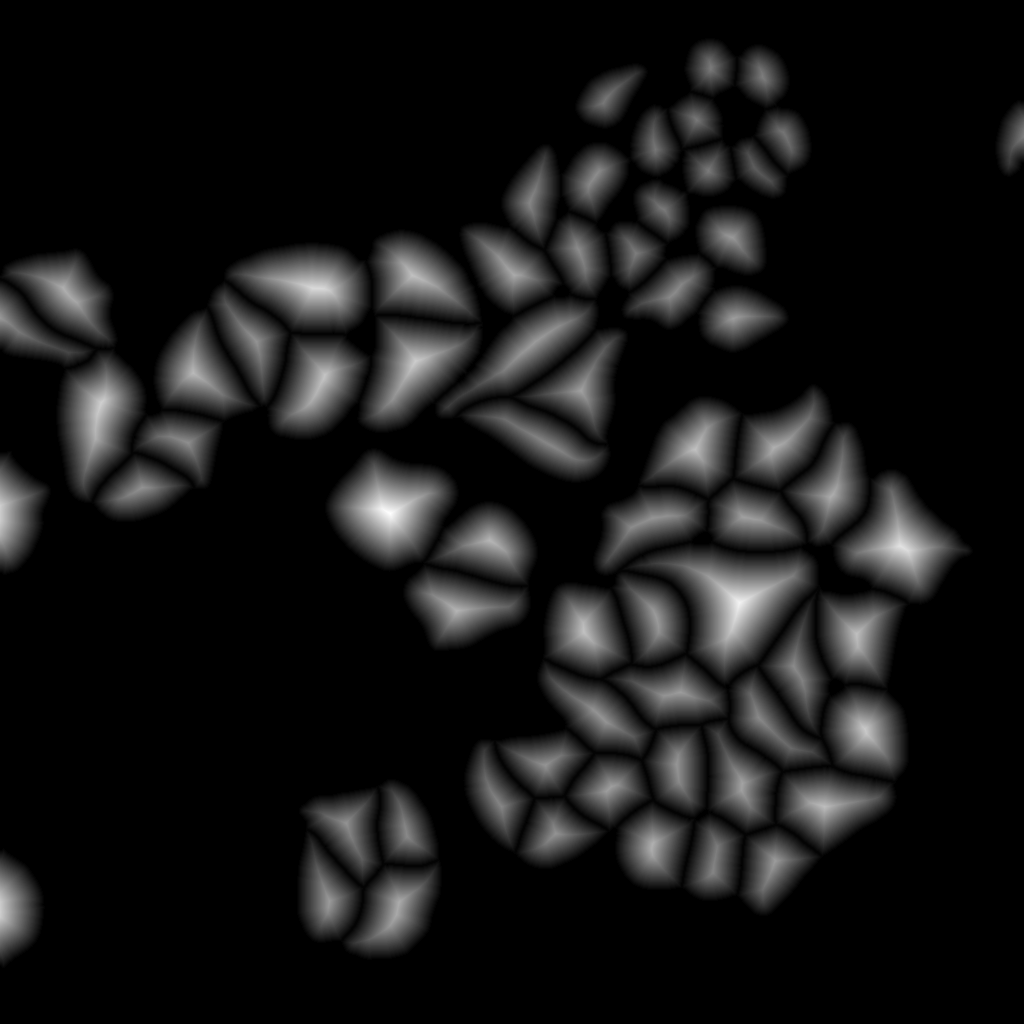}
   \caption{}
   \label{fig:cell_topology}
\end{subfigure}

\caption{(a) Fluorescent CellMask\texttrademark image. (b) Corresponding cell instance segmentation image generated by Cellpose. (c) Cell semantic segmentation image generated from Cellpose output. (d) Distance map generated from Cellpose output.}
\label{fig:Cellposecell} 
\end{figure}

Again, we used Cellpose \cite{cellpose} with manual correction to generate this instance segmentation ground truth images from associated CellMask\texttrademark images (Fig.\ref{fig:cy5_bis}, Fig.\ref{fig:cell_instance}). 

As for nuclear segmentation, we used training sets of different sizes $N \in \{1, 10, 50, 80\}$ of dimension (1024, 1024) and evaluated the accuracy for each of them. Testing is always performed on the same 17 images. 5 images of dimension (512, 512) were randomly cropped from each initial image.

To tackle the issue of instance segmentation, we implemented a model predicting both a cell semantic segmentation image (Fig.\ref{fig:cell_semantic}) and a distance map, i.e. an image where pixels values get higher as they are closer to the center of the cell, the background remaining black (Fig.\ref{fig:cell_topology}), as proposed in \cite{Naylor2017}, \cite{Naylor2019}.

\begin{figure}
\centering
\begin{subfigure}[b]{0.8\textwidth}
   \includegraphics[width=1\linewidth]{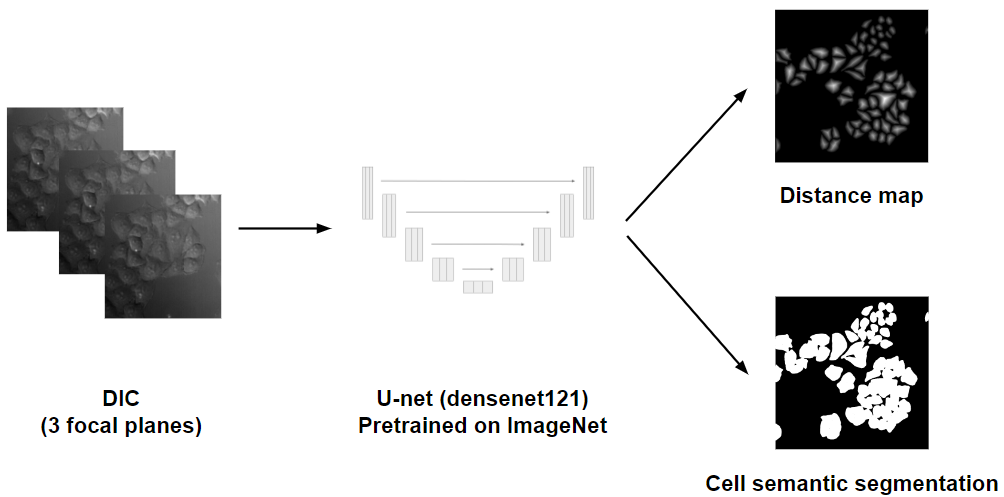}
   \caption{}
   \label{fig:cell_steroids}
\end{subfigure}

\begin{subfigure}[b]{0.8\textwidth}
   \includegraphics[width=1\linewidth]{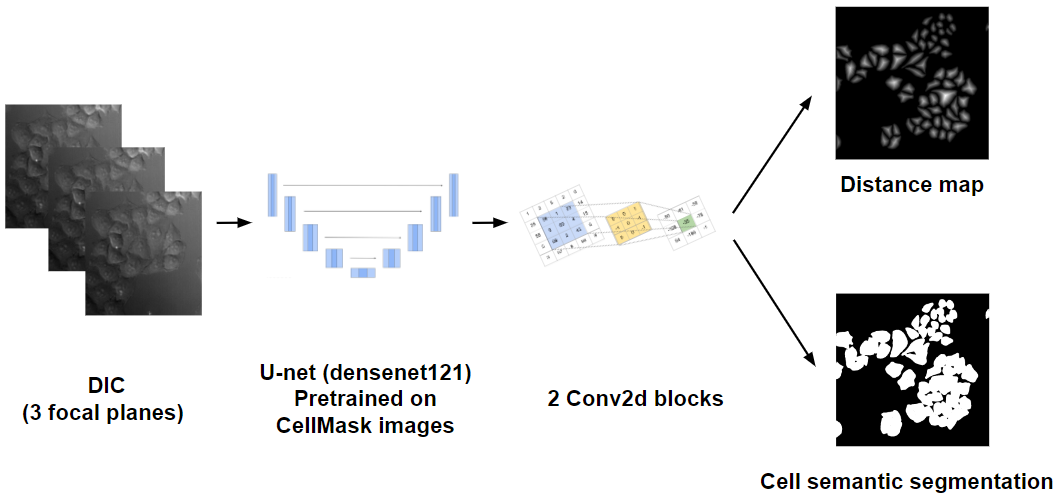}
   \caption{}
   \label{fig:cell_pretrained}
\end{subfigure}

\caption {Models compared to predict cell instance segmentation: (a) U-net 'on steroids' which has not been trained on CellMask\texttrademark images. (b) U-net model pretrained on CellMask\texttrademark images.}
\end{figure}

Like in the previous section we compare two models to investigate whether transfer learning from an ISL model can significantly improve the accuracy of our segmentation. The first model is the U-net 'on steroids', outputting 2 channels (Fig.\ref{fig:cell_steroids}). The second model has the same U-net architecture but is pretrained on CellMask\texttrademark images, thus outputting only 1 channel. Hence we add two Conv2d layers at the end to upscale to 2 channels (Fig.\ref{fig:cell_pretrained}). 

We did not use any data augmentation. All training details are reported in Supplementary Table 2. Both models use $\mbox{CombinedLoss}_{\alpha}$, presented in equation \eqref{loss_sum}, as loss function. MSELoss stands for the usual Mean Square Error, while BCEWithLogitsLoss combines a sigmoid layer with the Binary Cross Entropy loss. $y$ represents the output of our model, with the two channels $y_d$ and $y_s$ standing for the distance and semantic segmentation image, respectively. The factor $\alpha$ is used to balance the weights of the different losses during training. It has been set as $\alpha = 2000$, 2000 being the initial ratio between MSELoss and BCEWithLogitsLoss. This has been inspired by the loss function used in Cellpose \cite{cellpose}, which also uses a loss function computed as the sum of two loss functions, one for each output channel.

\begin{equation}
\begin{aligned}
    \mbox{CombinedLoss}_{\alpha}(y) & =\mbox{CombinedLoss}_{\alpha}((y_d, y_s))\\
         & =\mbox{MSELoss}(y_d) + \alpha \cdot \mbox{BCEWithLogitsLoss}(y_s)
\end{aligned}
\label{loss_sum}
\end{equation}

\begin{figure}
\centering
\includegraphics[width=0.8\linewidth]{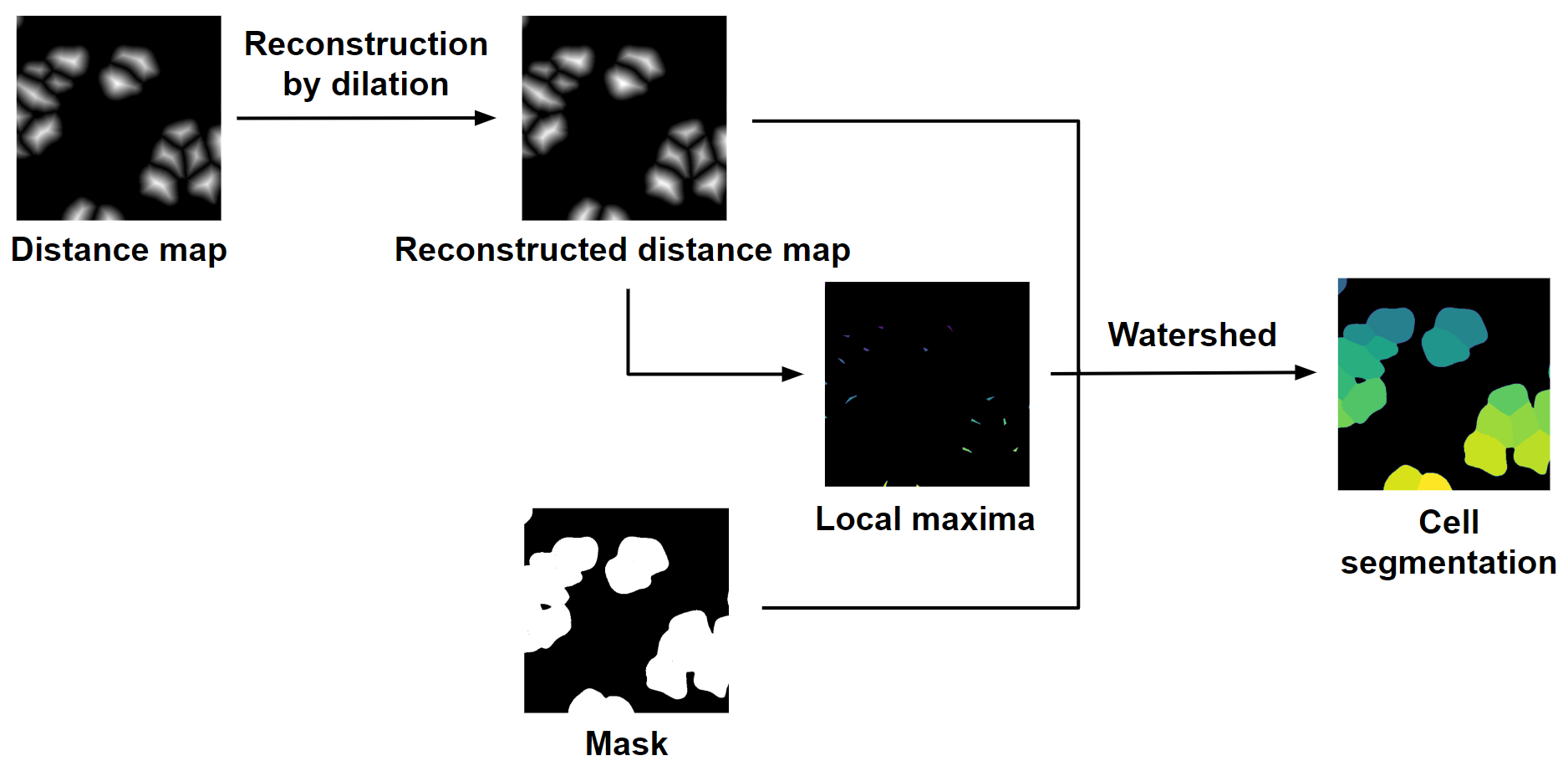}
\caption{Pipeline to get instance segmentation image from both distance map and semantic segmentation image. H-maxima transform followed by watershed algorithm enable to segment cells instance-wise.}
\label{fig:watershed} 
\end{figure}

Finally, we apply a post-processing step to get the final results. For this, we apply the h-maxima transformation of the predicted distance map, with $h=10$. The h-maxima transformation is defined as the reconstruction by dilation of $f-h$ under $f$: $HMAX_h(f) = R^{\delta}_f(f-h)$, and removes insignificant local maxima. $f$ stands for the initial image, which is in our case the reconstructed distance map displayed in Fig.\ref{fig:watershed}. $h$ stands for the minimum local contrast for a local maximum to be kept; otherwise it will be removed. Each local maximum represents a cell.

The local maxima of $HMAX$ then serve as seed for the watershed algorithm, which splits semantic segmentation result into individual regions, one for each maximum of $HMAX$. 
This leads to an instance segmentation image, such as the one presented in Fig.\ref{fig:watershed}.

\section{Results}

\subsection{Evaluation metrics}

The metric used to evaluate DAPI and CellMask\texttrademark prediction performance is the Pearson Correlation Coefficient (PCC, equation \eqref{PCC}). PCC is defined as the covariance of two variables divided by the product of their standard deviations. In the equation, $x$ and $y$ are two images to compare, and $\bar{x}$ is the average of $x$. 

\begin{equation}
\textup{PCC}(x,y)=\frac{\sum\limits_{i=0}^{n}{(x_i-\bar{x})(y_i-\bar{y})}}{\sqrt{\sum\limits_{i=0}^{n}{(x_i-\bar{x})^2}}\sqrt{\sum\limits_{i=0}^{n}{(y_i-\bar{y})^2}}}
\label{PCC}
\end{equation}

To evaluate nucleus semantic segmentation, we use the Jaccard index (equation \eqref{IoU}). The Jaccard index, or Intersection Over Union (IoU), is a very popular metric in segmentation, as it equally penalizes both False Positive and False Negative pixels. A perfect segmentation would lead to IoU of 1, while IoU corresponds to an entirely missed object (no intersection). 

\begin{equation}
\textup{IoU}(x, y) = \left\{\begin{matrix}
1 & \textup{ if } x \cup  y = 0 \\ 
\frac{x \cap  y}{x \cup  y} & \textup{ otherwise}
\end{matrix}\right.
\label{IoU}
\end{equation}

While the IoU is perfectly suitable to make pixel-wise comparisons for semantic segmentation, the performance of instance segmentation needs to incorporate an object-wise comparison that does not only penalize wrong pixel decisions, but also fused or split objects. For this, we choose to use the Mean Average Precision (mAP), which is a popular metric for instance segmentation evaluation. For this, a connected component from the ground truth is matched with a connected component from the segmentation result, if the IoU of the two components is above a given threshold. In this case, the object is considered as a TP. Unmatched connected components from the ground truth and the segmentation result are considered as FN and FP, respectively. 
Thus, given an IoU threshold one can compute the precision as defined in equation \eqref{precision}. 

\begin{equation}
\textup{Precision} = \frac{\textup{TP}}{\textup{TP} + \textup{FP} + \textup{FN}} 
\label{precision}
\end{equation}

Precision is computed for all 10 IoU thresholds in $\{0.5 + i \times  0.05, i \in [\![0, 9 ]\!] \}$. The final result is the mean of these 10 values, hence called mean AP, or mAP.

\subsection{Nucleus Segmentation}

DAPI prediction yields very good results, with a PCC of 0.95$\pm$0.08.

\begin{figure}
\centering
\begin{subfigure}[b]{0.4\textwidth}
    \centering
   \includegraphics[width=1\linewidth]{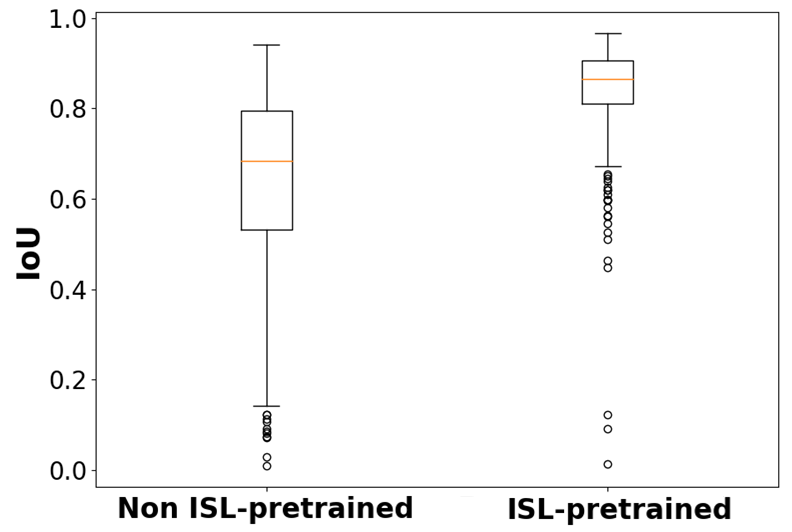}
   \caption{}
   \label{fig:box_plot_nuclei} 
\end{subfigure}%
~
\begin{subfigure}[b]{0.55\textwidth}
\centering
   \includegraphics[width=1\linewidth]{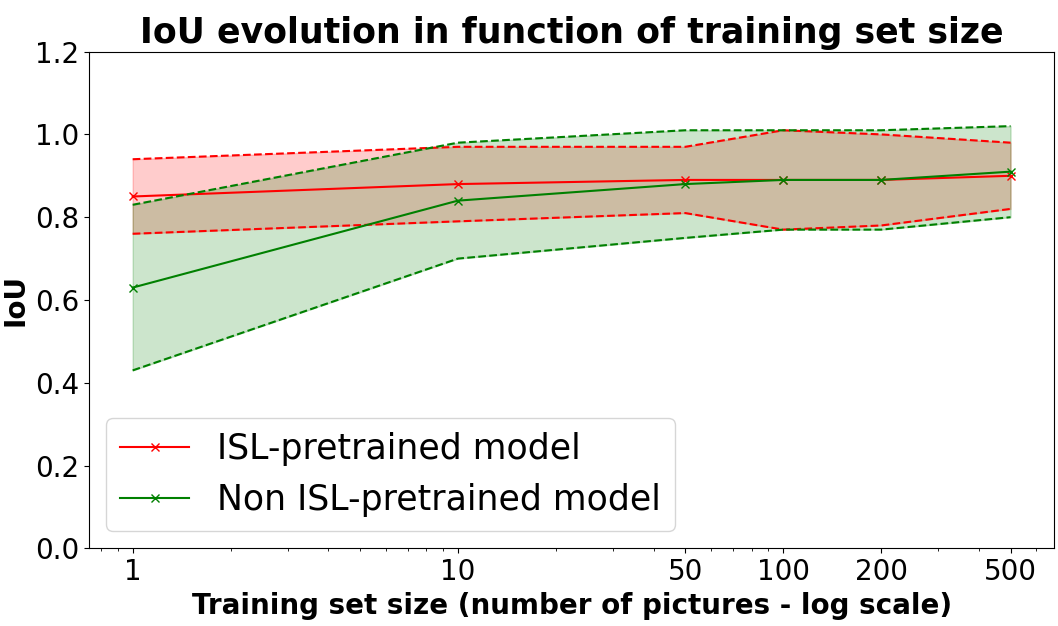}
   \caption{}
   \label{fig:diff_ts_nuclei}
\end{subfigure}

\caption {Nucleus segmentation results. (a) Intersection Over Union (IoU) score for non ISL-pretrained and ISL-pretrained models, after training on 1 image. (b) Evolution of IoU average score for both models for different training set sizes.}
\end{figure}

Using the Jaccard index (or IoU) as metric, the U-net 'on steroids' gives 0.64$\pm$0.2 after training on 1 single image. In comparison, the model pretrained on DAPI reaches 0.84$\pm$0.1, improving the previous score by 31.3\% (Fig.\ref{fig:box_plot_nuclei}). This improvement decreases as the size of the training set increases, being 4.8\% (respectively 1.1\%, 0.0\%, 0.0\%, -1.1\%) after training on 10 (respectively 50, 100, 200, 500) (Fig.\ref{fig:diff_ts_nuclei}).  

\begin{figure}
\centering
\includegraphics[width=1\linewidth]{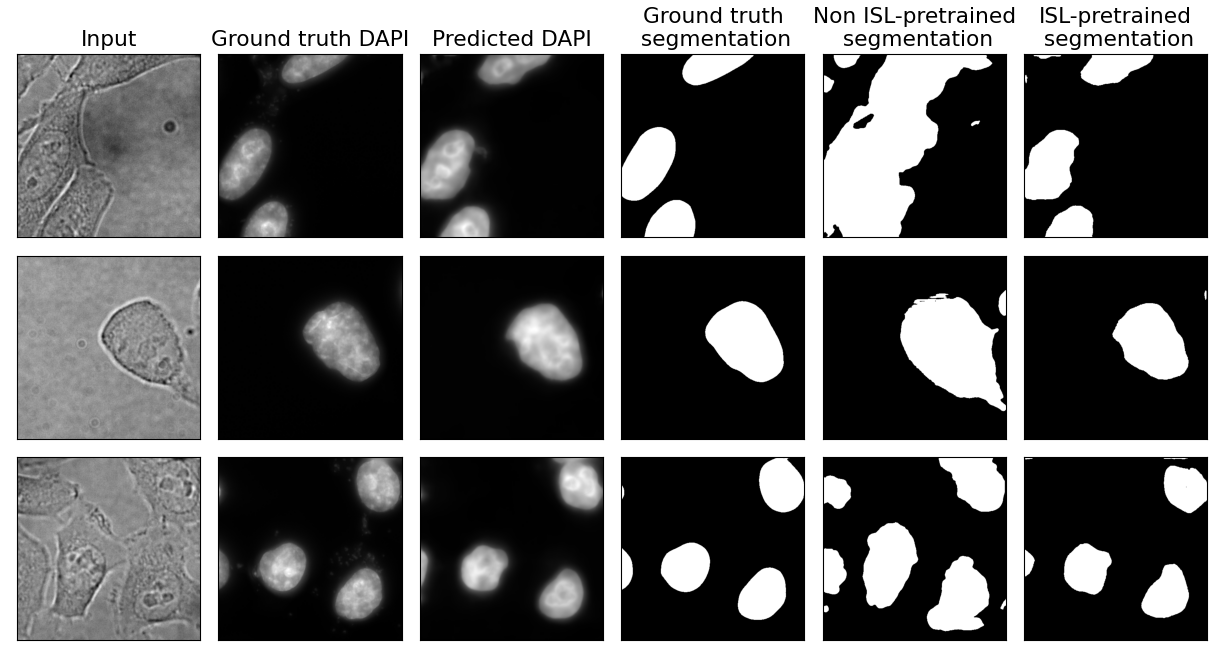}
\caption{Input bright-field images, DAPI images, DAPI predictions generated by U-net 'on steroids', ground truth instance segmentation generated by Cellpose, non ISL-pretrained U-net 'on steroids' segmentation prediction, ISL-pretrained U-net 'on steroids' segmentation prediction. Segmentation is performed after training on 1 image for both models.}
\label{fig:nucleiprediction} 
\end{figure}

Results from both models trained on 1 single image are displayed in Fig.\ref{fig:nucleiprediction}.

\subsection{Cell Segmentation}

CellMask\texttrademark prediction also yields very good results, with a PCC of 0.97$\pm$0.02.

\begin{figure}
\centering
\begin{subfigure}[b]{0.40\textwidth}
    \centering
   \includegraphics[width=1\linewidth]{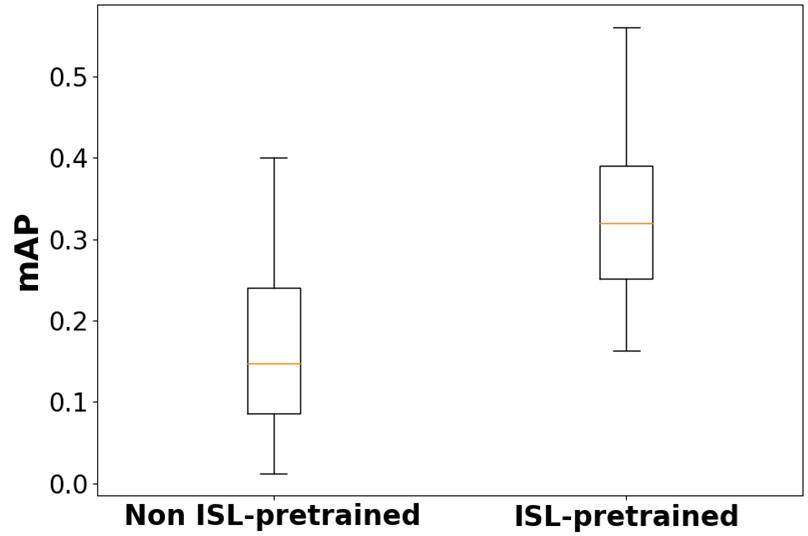}
   \caption{}
   \label{fig:box_plot_cells} 
\end{subfigure}%
~
\begin{subfigure}[b]{0.55\textwidth}
\centering
   \includegraphics[width=1\linewidth]{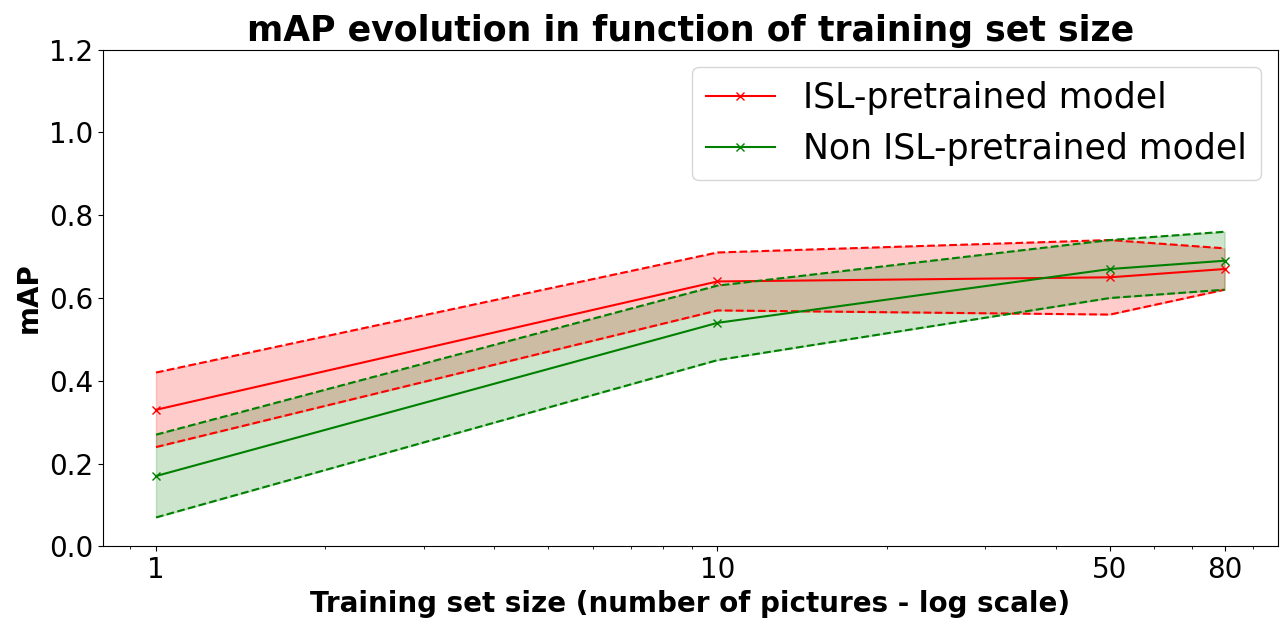}
   \caption{}
   \label{fig:diff_ts_cells}
\end{subfigure}

\caption {Cell segmentation results. (a) mAP score for non ISL-pretrained and ISL-pretrained models, after training on 1 image. (b) Evolution of mAP average score for both models for different training set sizes.}
\end{figure}

Using mAP as metric, the U-net 'on steroids' gives 0.17$\pm$0.1 after training on 1 single image. In comparison, the model pretrained on CellMask\texttrademark reaches 0.33$\pm$0.09, improving the previous score by 94.1\% (Fig.\ref{fig:box_plot_cells}). As in the previous section this improvement decreases as the size of the training set increases, being 18.5\% (respectively -3.0\%, -2.9\%) after training on 10 (respectively 50, 80) (Fig.\ref{fig:diff_ts_cells}).  

\begin{figure}
\centering
\includegraphics[width=1\linewidth]{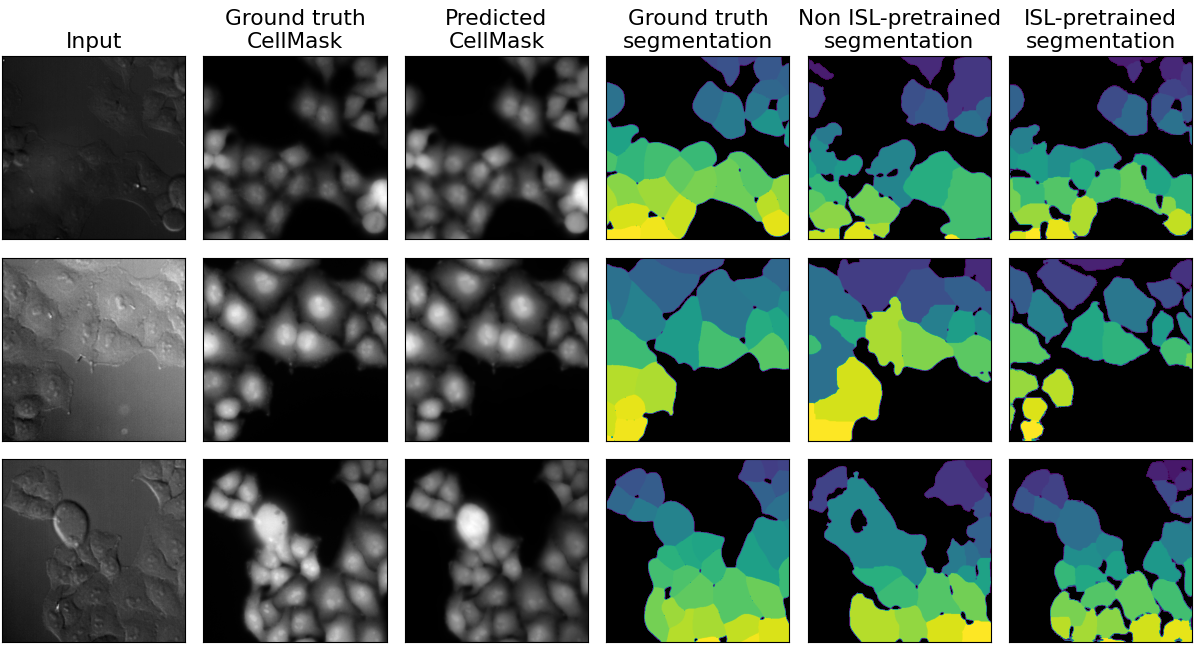}
\caption{Input DIC images, CellMask\texttrademark images, CellMask\texttrademark predictions generated by U-net 'on steroids', ground truth instance segmentation generated by Cellpose, non ISL-pretrained U-net 'on steroids' segmentation prediction, ISL-pretrained U-net 'on steroids' segmentation prediction. Segmentation is performed after training on 1 image for both models.}
\label{fig:cellsprediction} 
\end{figure}

Results from both models trained on 1 single image are displayed in Fig.\ref{fig:cellsprediction}.

\section{Discussion}

The results presented in the previous sections show that pretraining with In Silico Labeling as pretext task significantly improves the performance of a segmentation model trained on a very small data set. Indeed, the accuracy raises by 31.3\% and 94.1\% for nucleus semantic segmentation and cell instance segmentation, respectively, after training on 1 single image, using a model pretrained in an ISL setting.

The fact that pretraining on DAPI images helps to generate a nucleus semantic segmentation was actually expected since the two outputs (DAPI and binary segmentation maps) are very close to each other. On the other hand, cell instance segmentation is a much more complex problem, and our results clearly indicate that also in this situation, pretraining with fluorescent marker prediction as a pretext task significantly improves segmentation accuracy for small datasets. We also observe that transfer learning is useful if we work on a very small data set (1 to 10 images), but that for both nucleus and cytoplasmic segmentation, the accuracy difference disappears if the models are trained on more than 10 images. This being said, if one has access to fluorescent images, it makes sense to use our proposed method to pretrain the network. 

From a practical point of view, this idea provides an interesting alternative to manual annotation, in particular in the context of High Content Screening, where it is fairly easy to generate large amounts of data that contain both label-free and fluorescently labeled microscopy images. In this case, we can train efficient models for fluorescence prediction, and use these models in a pre-training scheme to reduce the manual annotation burden. Finally, we showed here that this pre-training scheme is effective for segmentation of nuclei and cells, but we also believe that this could be effective for any other type of cell structures as soon as you can get the associated fluorescent images available. Furthermore, it will be interesting to investigate to which extent the pre-training scheme provides good starting points for generalist networks, applicable to a wide variety of modalities.

\section{Conclusion}

In this paper, we demonstrated that pretraining on the prediction of relevant fluorescent markers can be very useful to segment nuclei or cells. We showed that a model trained to predict some fluorescent structures from label-free microscopy can learn to segment these structures from a very small data set, down to 1 single image. We believe that this can be of great help for applications where fluorescent data are easily available, if one wants to avoid tedious manual annotation to build large ground truth datasets for the training of neural networks. With only a few images, it is possible to fine-tune a pretrained model achieving performances matching those obtained by ImageNet-pretrained state-of-the-art networks fine-tuned on a much larger set of images. Our pre-training scheme can thus help biologists to save time and money without sacrificing any accuracy.

\subsubsection{Code availability}
Code (pre-processing, training and testing, post-processing pipelines), is available at \href{https://github.com/15bonte/isl_segmentation}{https://github.com/15bonte/isl\_segmentation}.

\subsubsection{Acknowledgments}
This work has been supported by the French government under management of Agence Nationale de la Recherche (ANR) as part of the “Investissements d’avenir” program, reference ANR-19-P3IA-0001 (PRAIRIE 3IA Institute), the Q-Life funded project CYTODEEP (ANR-17-CONV-0005) and the ANR project TRANSFACT (ANR-19-CE12-0007). Furthermore, we also acknowledge support by France-BioImaging (ANR-10-INBS-04). 

\pagebreak

\printbibliography

\pagebreak

\section*{Supplementary material}

\begin{table}[!ht]
\centering
\resizebox{!}{0.75\textheight}{%
\rotatebox{90}{
\begin{tabular}{ ccccccccccc }
 \hline
 \multirow{ 2}{*}{Input} & \multirow{ 2}{*}{Output} & \multirow{ 2}{*}{Pretrain} & \multirow{ 2}{*}{Images} & \multirow{ 2}{*}{Crops} & Training  & \multirow{ 2}{*}{Epochs} & Learning  & \multirow{ 2}{*}{Loss} & Training  & \multirow{ 2}{*}{Evaluation} \\
  &  &  &  &  &  samples &  &  rate &  &  time &  \\
 \hline
BF + PC & DAPI &  ImageNet & 384 & 1920 & 192 000 & 1000 & 0.1 & L1 & $\sim$2d & PCC \\
\hline
BF + PC & Nuclei & ImageNet & 1 & 5 & 5 000 & 5 000 & 0.01 & Jaccard & $\sim$2h & IoU \\
BF + PC & Nuclei & ImageNet & 2 & 10 & 5 000 & 5 000 & 0.01 & Jaccard & $\sim$2h & IoU \\
BF + PC & Nuclei & ImageNet & 10 & 50 & 5 000 & 1 000 & 0.01 & Jaccard & $\sim$2h & IoU \\
BF + PC & Nuclei & ImageNet & 50 & 250 & 10 000 & 400 & 0.01 & Jaccard & $\sim$2h & IoU \\
BF + PC & Nuclei & ImageNet & 100 & 500 & 10 000 & 200 & 0.01 & Jaccard & $\sim$2h & IoU \\
\hline
BF + PC & Nuclei & ISL (DAPI) & 1 & 5 & 20 000 & 20 000 & 0.01 & Jaccard & $\sim$3h & IoU \\
BF + PC & Nuclei & ISL (DAPI) & 2 & 10 & 20 000 & 20 000 & 0.01 & Jaccard & $\sim$3h & IoU \\
BF + PC & Nuclei & ISL (DAPI) & 10 & 50 & 20 000 & 4 000 & 0.01 & Jaccard & $\sim$3h & IoU \\
BF + PC & Nuclei & ISL (DAPI) & 50 & 250 & 20 000 & 800 & 0.01 & Jaccard & $\sim$3h & IoU \\
BF + PC & Nuclei & ISL (DAPI) & 100 & 500 & 20 000 & 400 & 0.01 & Jaccard & $\sim$3h & IoU \\
\hline
\end{tabular}}
}%
\caption{Training details for DAPI and nucleus segmentation models. Segmentation models were supposed to be trained on 20 000 training samples, yet for non ISL-pretrained models the training was stopped earlier as the model was not learning anything after some point. All models have the same U-net densenet121 architecture. They were trained with an ADAM optimizer on GPU hardware. BF stands for bright-field and PC for phase contrast.}
\label{table:nucleus_models}
\end{table}

\begin{table}[!ht]
\centering
\resizebox{!}{0.9\textheight}{%
\rotatebox{90}{
\begin{tabular}{ ccccccccccc }
 \hline
 \multirow{ 2}{*}{Input} & \multirow{ 2}{*}{Output} & \multirow{ 2}{*}{Pretrain} & \multirow{ 2}{*}{Images} & \multirow{ 2}{*}{Crops} & Training  & \multirow{ 2}{*}{Epochs} & Learning  & \multirow{ 2}{*}{Loss} & Training  & \multirow{ 2}{*}{Evaluation} \\
  &  &  &  &  &  samples &  &  rate &  &  time &  \\
 \hline 
DIC & CellMask\texttrademark &  ImageNet & 65 & 325 & 132 000 & 4000 & 0.1 & L1 & $\sim$1d & PCC \\
\hline
DIC & Distance + Mask & ImageNet & 1 & 5 & 10 000 & 10 000 & 0.01 & $\mbox{CombinedLoss}_{\alpha}$ & $\sim$1h30 & mAP \\
DIC & Distance + Mask & ImageNet & 2 & 10 & 10 000 & 10 000 & 0.01 & $\mbox{CombinedLoss}_{\alpha}$ & $\sim$1h30 & mAP \\
DIC & Distance + Mask & ImageNet & 10 & 50 & 10 000 & 2 000 & 0.01 & $\mbox{CombinedLoss}_{\alpha}$ & $\sim$1h30 & mAP \\
DIC & Distance + Mask & ImageNet & 50 & 250 & 10 000 & 400 & 0.01 & $\mbox{CombinedLoss}_{\alpha}$ & $\sim$1h30 & mAP \\
DIC & Distance + Mask & ImageNet & 80 & 400 & 10 000 & 250 & 0.01 & $\mbox{CombinedLoss}_{\alpha}$ & $\sim$1h30 & mAP \\
\hline
DIC & Distance + Mask & ISL (CellMask\texttrademark) & 1 & 5 & 10 000 & 10 000 & 0.01 & $\mbox{CombinedLoss}_{\alpha}$ & $\sim$2h & mAP \\
DIC & Distance + Mask & ISL (CellMask\texttrademark) & 2 & 10 & 10 000 & 10 000 & 0.01 & $\mbox{CombinedLoss}_{\alpha}$ & $\sim$2h & mAP \\
DIC & Distance + Mask & ISL (CellMask\texttrademark) & 10 & 50 & 10 000 & 2 000 & 0.01 & $\mbox{CombinedLoss}_{\alpha}$ & $\sim$2h & mAP \\
DIC & Distance + Mask & ISL (CellMask\texttrademark) & 50 & 250 & 10 000 & 400 & 0.01 & $\mbox{CombinedLoss}_{\alpha}$ & $\sim$2h & mAP \\
DIC & Distance + Mask & ISL (CellMask\texttrademark) & 80 & 400 & 10 000 & 250 & 0.01 & $\mbox{CombinedLoss}_{\alpha}$ & $\sim$2h & mAP \\
\hline
\end{tabular}}
}%
\caption{Training details for CellMask\texttrademark and cell segmentation models. All models have the same U-net densenet121 architecture. They were trained with an ADAM optimizer on GPU hardware.}
\label{table:cell_models}
\end{table}

\end{document}